\documentclass{article}

\usepackage[final]{neurips_2024_ml4ps}

\usepackage[utf8]{inputenc} 
\usepackage[T1]{fontenc}    
\usepackage{hyperref}       
\usepackage{url}            
\usepackage{booktabs}       
\usepackage{amsfonts}       
\usepackage{nicefrac}       
\usepackage{microtype}      
\usepackage{xcolor}         

\usepackage{svg}
\usepackage{amsmath}
\usepackage{graphicx}

\usepackage{float}

\title{MRI Parameter Mapping via Gaussian Mixture VAE: Breaking the Assumption of Independent Pixels}

\author{
    Moucheng Xu $^{1,2,6}$\thanks{Work completed at UCL, Moucheng has now moved to Medtronic.}
    \And Yukun Zhou $^{1,6}$
    \And Tobias Goodwin-Allcock $^{1,6}$
    \And Kimia Firoozabadi $^{5}$
    \And Joseph Jacob $^{1,2,3}$
    \And Daniel C. Alexander $^{1,4}$ 
    \And Paddy J. Slator $^{1,4,7,8}$ \\ \\
    $^1$UCL Hawkes Institute, UK $^2$UCL Satsuma Lab, UK $^3$UCL Respiratory, UK \\
    $^4$UCL Department of Computer Science, UK $^5$UCL Medical School, UK\\
    $^6$UCL Department of Medical Physics and Biomedical Engineering, UK \\
    $^7$Cardiff University, School of Computer Science and Informatics, UK \\
    $^8$Cardiff University Brain Research Imaging Center, UK \\
    \texttt{moucheng.xu.18@alumni.ucl.ac.uk} \\
    \texttt{slatorp@cardiff.ac.uk}
    }

\begin{document}

\maketitle

\begin{abstract}
We introduce and demonstrate a new paradigm for quantitative parameter mapping in MRI.
Parameter mapping techniques, such as diffusion MRI and quantitative MRI, have the potential to robustly and repeatably measure biologically-relevant tissue maps that strongly relate to underlying microstructure.
Quantitative maps are calculated by fitting a model to multiple images, e.g. with least-squares or machine learning. However, the overwhelming majority of model fitting techniques assume that each voxel is independent, ignoring any co-dependencies in the data. This makes model fitting sensitive to voxelwise measurement noise,  hampering reliability and repeatability.
We propose a self-supervised deep variational approach that breaks the assumption of independent pixels, leveraging redundancies in the data to effectively perform data-driven regularisation of quantitative maps.  
We demonstrate that our approach outperforms current model fitting techniques in dMRI simulations and real data. Especially with a Gaussian mixture prior, our model enables sharper quantitative maps, revealing finer anatomical details that are not presented in the baselines. Our approach can hence support the clinical adoption of parameter mapping methods such as dMRI and qMRI. Our code is available at \href{https://github.com/moucheng2017/MRI-GMM-VAE}{https://github.com/moucheng2017/MRI-GMM-VAE}. 
\end{abstract}

\section{Introduction}

Multiple MRI techniques can produce quantitative maps of biophysical, chemical and physiological tissue properties.
Such \textit{quantitative parameter mapping} techniques include diffusion MRI (dMRI) and quantitative MRI (qMRI).
dMRI and qMRI use an essentially identical approach; by acquiring multiple images then fitting a model to the images, intrinsic values of the relevant tissue properties in each voxel can be estimated. 
In dMRI the images have different diffusion weightings and directions, and model fitting enables estimation of microstructural parameters, such as diffusivity and kurtosis.
In qMRI acquisition parameters such as echo time or inversion time are varied, and model fits produce maps of chemical  tissue properties, such as T1 and T2.


In the vast majority of applications, such models are fit to the data using non-linear least squares techniques. 
Machine learning model fitting is emerging as a attractive alternative technique.
Supervised learning has been demonstrated for a range of models \cite{palomboSANDICompartmentbasedModel2020}, but the  distribution of  parameters in the training dataset introduces biases \cite{gyoriTrainingDataDistribution2022,epsteinChoiceTrainingLabel2022}. 
Self-supervised learning has the potential to address this, but has only been demonstrated in a few models thus far, most prominently the intravoxel incoherent motion (IVIM) model \cite{barbieriDeepLearningHow2020}. 
Hybrid approaches are also emerging that merge the benefits of supervised and self-supervised learning \cite{epsteinChoiceTrainingLabel2022}.

However, whilst these approaches offer improvements, the current generation of parameter mapping techniques fail to capitalize on the extensive inherent redundancies in the data.
Specifically, the overwhelming majority of techniques fit models to each voxel separately, effectively assuming that each voxel is independent.
This leads to high sensitivity to voxelwise measurement noise, which negatively affects the quality of derived parameter maps.
Bayesian hierarchical modelling has been proposed as an approach that breaks these assumptions, but requires slow Markov chain Monte Carlo inference \cite{ortonImprovedIntravoxelIncoherent2014}. 
Convolutional neural networks (CNNs)  \cite{vasylechkoSelfsupervisedIVIMDWI2021} have been demonstrated, but only learn spatial redundancies. Transformers have also been demonstrated \cite{Zheng2023} but rely on gold standard training data.
One technique assumes a set of underlying tissue components to regularise quantitative maps \cite{slatorDataDrivenMultiContrastSpectral2021} in a data-driven and unsupervised way, but at the expense of voxelwise parameter estimates. 

In this paper, we demonstrate a deep learning approach that breaks the paradigm of independent voxels.
Analogously to recent approaches \cite{slatorDataDrivenMultiContrastSpectral2021}, we seek a lower dimensional representation of the data to parameterise date redundancies.
We therefore adapt ideas from the clustering literature \cite{manduchiDeepVariationalApproach2022a} to derive a deep variational autoencoder for quantitative parameter mapping. We show that our approach yields improved parameter estimates and maps in simulated and real data.

\section{Methods}
\textbf{Problem Setting} We assume a dMRI dataset and model throughout the paper. We assume an observed discrete series of, $T$, dMRI images, $ \mathcal{S} = S_{(1)}, S_{(2)}, ..., S_{(T)}$, where $S_{(i)}$ is a diffusion weighted image (DWI) with height, $H$, width, $W$, and depth, $D$. The DWIs are defined $S_{(i)} = \{s_{(i)}^{(h,w,d)}; h \in [1,2,..,H], w \in [1,2,..,W], d \in [1,2,..,D] \}$, where $s_{(i)}^{(h,w,d)}$ is the signal at the voxel $(h,w,d)$ for the $i$th DWI. 
We aim to estimate the tissue properties, i.e. model parameters, $ X = \{ \textbf{x}^{(h,w,d)}; h \in [1,2,..,H], w \in [1,2,..,W], d \in [1,2,..,D] \}$ where the tissue properties for a voxel located at $(h,w,d)$ corresponds to the vector $\textbf{x}^{(h,w,d)}=(x^{(h,w,d)}_1,...,x^{(h,w,d)}_M)$ for the $M$ tissue properties. 
The model parameters are estimated with respect to the signal model.
Traditionally, each voxel would be estimated independently of all the other voxels, for example, the MLE at $(h_1, w_1, d_1)$ is calculated by finding the parameters that maximise the probability of seeing the observed data; i.e. what maximises $p((s_{(1)}^{(h_1, w_1, d_1)}, ..., s_{(T)}^{(h_1, w_1, d_1)})| \textbf{x}^{(h_1, w_1, d_1)}  )$. This is performed independently to the MLE at $(h_2, w_2, d_2)$. 

\textbf{Latent Variable Model} In this work, we propose to jointly model the distribution of the voxels together as $p_{\theta}(\mathcal{S})$ using a latent variable ($\textbf{z}$) model: $p_{\theta}(\mathcal{S}| \textbf{z}) p_{\theta}(z) dz$. Considering dMRI-specific format, with independence assumption for expression clarity (we note that despite this assumption our proposed model does not treat voxels as independent due to the shared latent space): $p_{\theta}(\mathcal{S}) \approx  \prod_{t=1}^{t=T}\prod_{d=1}^{d=D} \prod_{w=1}^{w=W} \prod_{h=1}^{h=H} p_\theta(s^{(h, w, d)}_{(t)}|\textbf{z}) p_\theta(\textbf{z})$. Our latent variable model enjoys two benefits. The first benefit is, by conditioning the data from each DWI from each voxel $S^{(h,w,d)}_{(i)}$ on $\textbf{z}$, we absorb all the arbitrary dependencies among voxels into $\textbf{z}$, a compact representation in a latent space. In the latent space which has lower dimension than the data space, voxels are clustered with their close voxels, therefore inter-voxels information must be captured. The second benefit is, the complicated underlying distribution of $p_\theta(\mathcal{S})$ can be learnt via learning a much simpler distribution $p_\theta(\textbf{z})$. 
However, the marginal distribution of $p_\theta(\textbf{z})$ is still intractable: $p_\theta(\textbf{z}|\mathcal{S}) = p_{\theta}(\mathcal{S}, \textbf{z}) / p_{\theta}(\mathcal{S})$ because the data density is unknown. To address this computational issue, we deploy a variational approach to approximate the posterior of $p_\theta(\textbf{z})$. We explore different implementations to obtain the posterior of $p(\textbf{z})$. 

\textbf{Univariate Gaussian prior} We start with a simple univariate Gaussian $\mathcal{N}(0, 1)$ for prior $q(\textbf{z})$ and our first implementation is called VAE-UniG. We first use the encoder to parameterize $\textbf{z}$ from input observed signals: $ \mu, Log(\sigma^2) = \theta_{Encoder}(\mathcal{S})$. Then, $p(\textbf{z} | \mathcal{S}, \theta_{Encoder}) \approx  \mathcal{N}(\mu, \sigma)$. We then use a decoder to map randomly drawn samples $\textbf{z} \sim p(\textbf{z})$ to physically-relevant dMRI model parameters ($X$) that are inputs to a closed-form dMRI model that reconstructs the MRI signal. We denote the closed-form physics decoding process as $\phi(.)$. We emphasise that $\phi(.)$ can be any dMRI (or qMRI) model. The complete decoding process is: $p(\mathcal{S}) = p(\phi(X | \textbf{z}, \theta_{Decoder}))$. The loss function is: $Log (p_\theta(\mathcal{S})) \geq \mathbb{E}_{z \sim p(\textbf{z})} [Log (p(\phi(\textbf{X} | \textbf{z}, \theta)))] - KL(p(\textbf{z}) || q(\textbf{z}))$. The likelihood of $Log p(\phi(\textbf{x} | \textbf{z}, \theta)))$ is measured as a mean squared error loss between estimated signals and raw input signals.

\textbf{Gaussian Mixture Prior} Our second implementation is VAE-GMM with enhanced expressivity by considering a prior as a mixture of univariate Gaussians. We add an extra latent variable $\textbf{y}$ for controlling the index of the Gaussian component. The prior of $\textbf{y}$ is chosen as a Categorical distribution. The probabilistic model is: $p_{\theta}(\mathcal{S}) =  \int_{\textbf{z}} \int_{\textbf{y}} p_\theta(\mathcal{S} | \textbf{z}) p (\textbf{z} | \textbf{y}) p(\textbf{y})  d \textbf{y} d\textbf{z}$. In implementation, we build our Gaussian mixture VAE (VAE-GMM) following a hierarchical order. We first need to parametrize the mixing coeffients of each Gaussian using Gumbel Softmax \cite{Maddison2016TheCD}: $c = Gumbel(\theta_{Encoder_{1st}}(\mathcal{S}))$. Where $c$ is a normalised vector indicating the weight for each Gaussian and the sum of the $c$ is 1. We then concatenate $c$ with input to parametrize the parameters of Gaussians: $\mu_k, Log(\sigma_k^2) = \theta_{Encoder_{2nd}}(\mathcal{S}, c)$. Also, $p(\textbf{z} | \mathcal{S}, \theta_{Encoder}) \approx  \mathcal{N}(\mu_k, \sigma_k)$. Where $k$ means that the mean and the variance are for the K-th Gaussian. We apply the same decoding process as in the last section. And the loss function now becomes: $Log (p_\theta(\mathcal{S})) \geq \mathbb{E}_{z \sim p(\textbf{y} | \textbf{z})p(\textbf{z} | x)} [Log (p(\phi(\textbf{X} | \textbf{z}, \theta)))] - KL(p(\textbf{z}, \textbf{y}) || q(\textbf{z}, \textbf{y}))$.

\textbf{MRI physics models} We test our approach on two dMRI models, the mean signal diffusion kurtosis imaging (MS-DKI)\cite{henriquesMicroscopicAnisotropyMisestimation2019} model and ball-stick model \cite{behrensCharacterizationPropagationUncertainty2003}. The normalised signal for MS-DKI \footnote{For MSDKI we implemented LSQ fitting with the dipy python package \cite{garyfallidisDipyLibraryAnalysis2014}.} is given by: $\phi(b) = \exp{(-bD + b^2 D^2 K / 6)}$, where $b$ is the b-value, $D$ is the diffusivity and $K$ the kurtosis. For ball-stick \footnote{For ball-stick we used the dmipy package \cite{fickDmipyToolboxDiffusion2019}.} the normalised signal is: $\phi(b, \mathbf{g}) = f \exp\left(-b D_{||} (\mathbf{g}.\mathbf{n})\right) + (1-f) \exp\left(-bD_{iso}\right)$, where $b$ is the b-value, $\mathbf{g}$ the gradient direction,  $f$ is the stick volume fraction, $D_{||}$ is the parallel diffusivity of the stick, and $D_{iso}$ is the ball isotropic diffusivity.

\begin{figure}[H]
\centering
\includegraphics[width=0.9\textwidth]{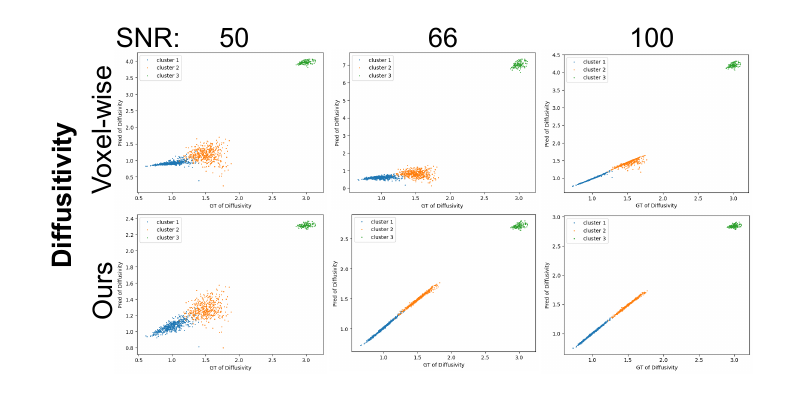}
\caption{Diffusivity results on simulated model using MSDKI, comparisons between self-supervised baseline and our VAE-UniG. X axis: ground truth of simulated diffusivity. Y axis: prediction of diffusivity. SNR: signal-noise-ratio. Blue: cluster 1. Orange: cluster 2. Green: cluster 3. Ours vastly outperforms the baseline in recovery of the ground truth. See the kurtosis results in the appendix Fig.\ref{fig:simulated_msdki_k}.}
\label{fig:simulated_msdki}
\end{figure}

\section{Results}
\textbf{Simulated Data} We use least squares fitting and voxel-wise self-supervised fitting (a 3-layer fully connected network \cite{limFittingDirectionalMicrostructure2022}) as baselines. We first test our method on a simulated MSDKI dataset with 3 clusters. We chose each cluster's diffusivity and kurtosis to mimic white matter, grey matter, and CSF; the mean ${D,K}$ values for each cluster were $\{1,1.5\}$, $\{1.5,1\}$, and $\{3,0\}$  respectively, with diffusivity in units of $\mu$m$^2$/ms. We simulated 10,000 voxels, with relative weightings of each cluster $\{0.5,0.4,1\}$. The specific ground truth parameter value was simulated from a Gaussian with the relevant mean $D$ and $K$ for that cluster, and variance $0.1$ for white matter and grey matter clusters, and $0.01$ for the CSF cluster. As shown in the Fig.~\ref{fig:simulated_msdki}, across different settings of signal-noise ratio (SNR), our model's predictions (row 2) are much better diagonally aligned, meaning that our latent variable model consistently outperform the voxel-wise self-supervised baseline (row 1).

\begin{figure}[H]
\centering
\includegraphics[width=\textwidth]{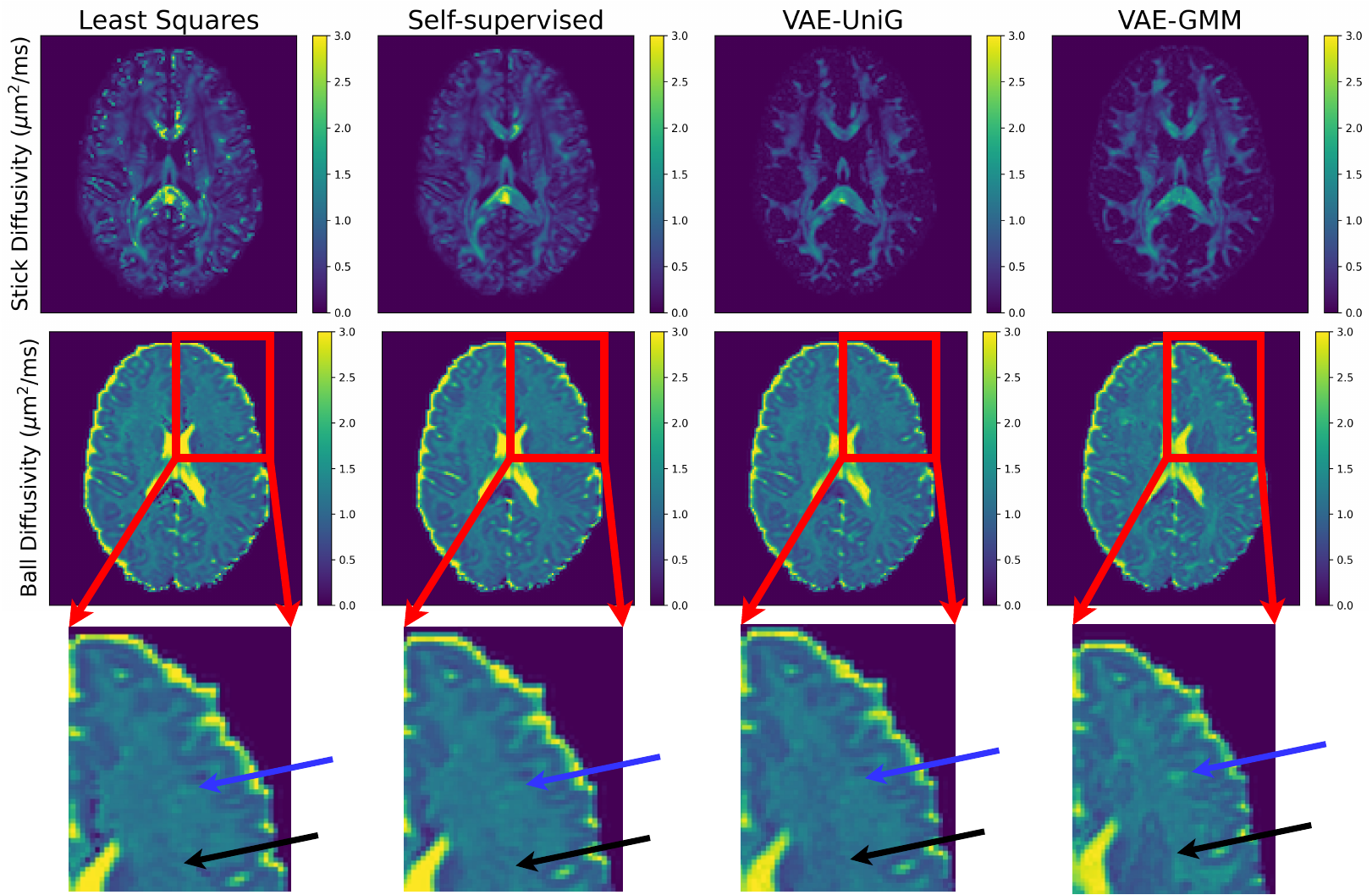}
\caption{LSQ, self-supervised, VAE-UniG, VAE-GMM ball-stick fits to HCP dMRI subject. Both our VAE models drastically reduce noises (1st Row) for sharper white matter visulisation. More importantly, our VAE-GMM reveals finer anatomical structures with clear details which were not seen in the baselines (see the arrows locations in 3rd Row, 4th Col).}
\label{fig:ball_stick_HCP}
\end{figure}

\begin{figure}[H]
\centering
\includegraphics[width=0.9\textwidth]{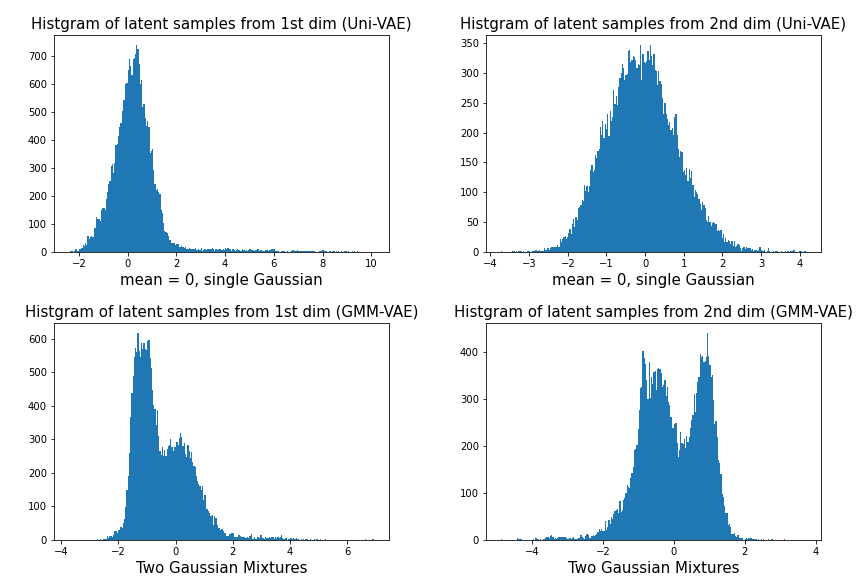}
\caption{Visualisation of the learnt posterior distributions of the latent variable $z$ training on HCP data. 1st row: VAE. 2nd row: GMM VAE.}
\label{fig:visual_z}
\end{figure}

\textbf{Real Data} We as well test our model on real data with ball-stick fit, with publicly-available dMRI data from HCP WU-Minn Consortium\cite{vanessenHumanConnectomeProject2012}. We used preprocessed\cite{glasserMinimalPreprocessingPipelines2013} data from a single subject from the 1200 Subjects Data Release (Release Date: Mar 01, 2017, available online at http://humanconnectome.org). As shown in the 1st row in Fig.~\ref{fig:ball_stick_HCP}, when predicting the diffusivity parameters, both of our models can drastically reduce the background noises in comparison with the baselines. Interestingly, our model VAE-GMM (4th column) can even discover new anatomical structures, as seen in highlighted areas in row3 in Fig.~\ref{fig:ball_stick_HCP}, showing potential clinical application promises. Evidently, our model VAE-GMM also successfully captured at least two mixing components of the latent variable, as shown in Fig.~\ref{fig:visual_z}.

\section{Conclusion}
We introduce a deep VAE model fitting method that exploits data redundancies to maximise information extraction in parameter mapping techniques like dMRI and qMRI. Our approach outperforms baseline methods in both simulated and real data using the ball-stick model, improving ground truth estimations and revealing new anatomical details. It can enhance existing acquisition sequences for better tissue maps or shorten scan times without loss of quality. However, limitations include variability in parameter maps and trade-offs between parameter accuracy. Future work will focus on generalising the method to handle arbitrary acquisition schemes.

\section{Acknowledgements}
MCX was supported by GSK (BIDS3000034123) and UCL Engineering Dean’s Prize. DCA is supported by UK EPSRC grants M020533, R006032, R014019, V034537, Wellcome Trust UNS113739, Wellcome Trust 221915/Z/20/Z. JJ is supported by Wellcome Trust Clinical Research Career Development Fellowship 227835/Z/23/Z. DCA, and JJ are supported by the NIHR UCLH Biomedical Research Centre, UK. This research was funded in whole or in part by the Wellcome Trust [209553/Z/17/Z]. 

\section{Impact Statement}
Medical physics plays a pivotal role in advancing physical sciences for the betterment of humanity. Among the many branches of this field, Magnetic Resonance Imaging (MRI) stands out as a critical area of study due to its wide-ranging applications in diagnostics, treatment planning, and medical research. MRI has revolutionized healthcare, enabling non-invasive imaging with detailed insights into the human body, making it indispensable for modern medicine. Our proposed method aims to further enhance MRI parameter mapping, improving both accuracy and efficiency. This has the potential to significantly impact the medical community by providing more reliable tools for diagnosis and research. Moreover, the method is versatile and can be adapted for use in a wide variety of parameter mapping applications beyond MRI, broadening its utility across multiple domains in medical physics and other scientific fields. By offering a solution that is both robust and adaptable, our work contributes to the broader goal of leveraging machine learning for physical sciences, providing meaningful societal benefits through improved medical imaging techniques and related applications.

\bibliographystyle{plainnat}
\bibliography{MyLibrary.bib}
\clearpage

\appendix
\section{The architectures of the neural networks}
\begin{figure}[H]
\centering
\includegraphics[width=0.9\textwidth]{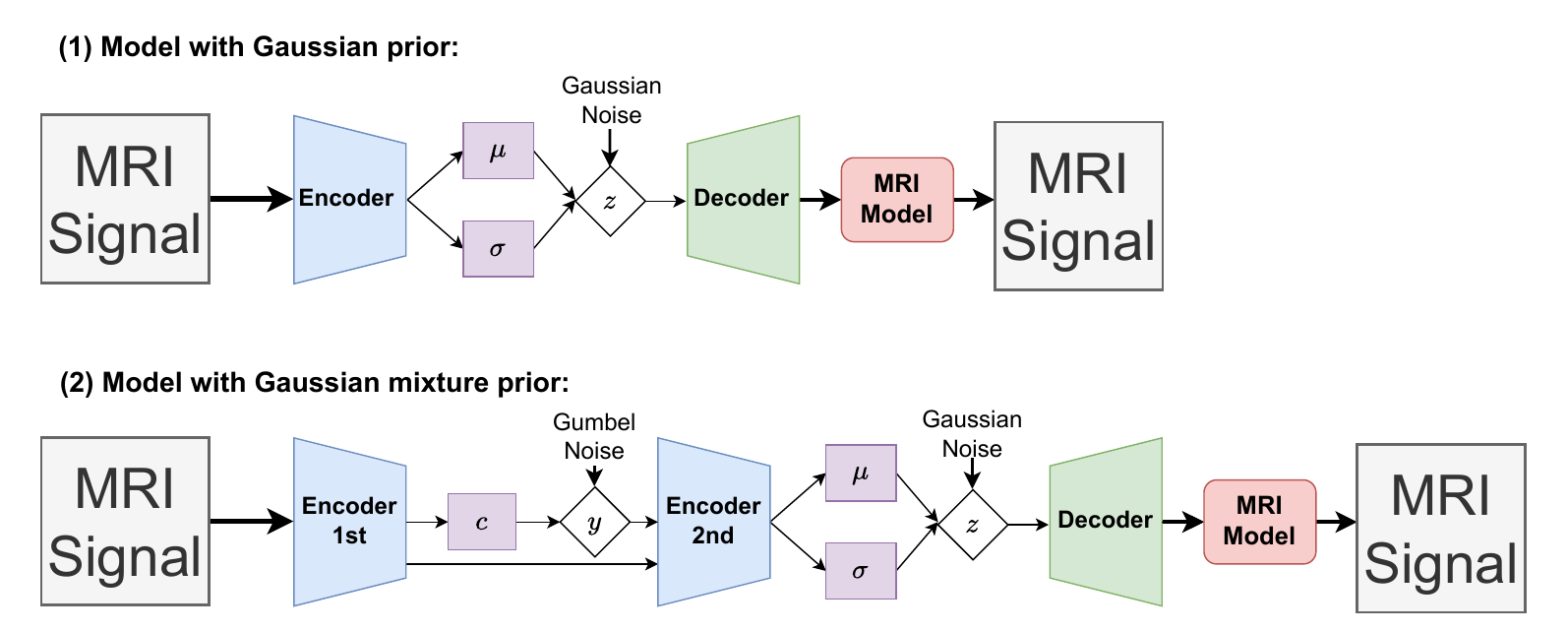}
\caption{Architectures of our model implementations. Row 1: with univariate prior. Row 2: with Gaussian mixture model prior. The encoder is three fully connected layers. The decoder is one fully connected layer.}
\label{fig:vae_arch}
\end{figure}

\section{Results on MSDKI MRI model}
\begin{figure}[H]
    \centering
    \begin{center}
        \includegraphics[width=\textwidth]{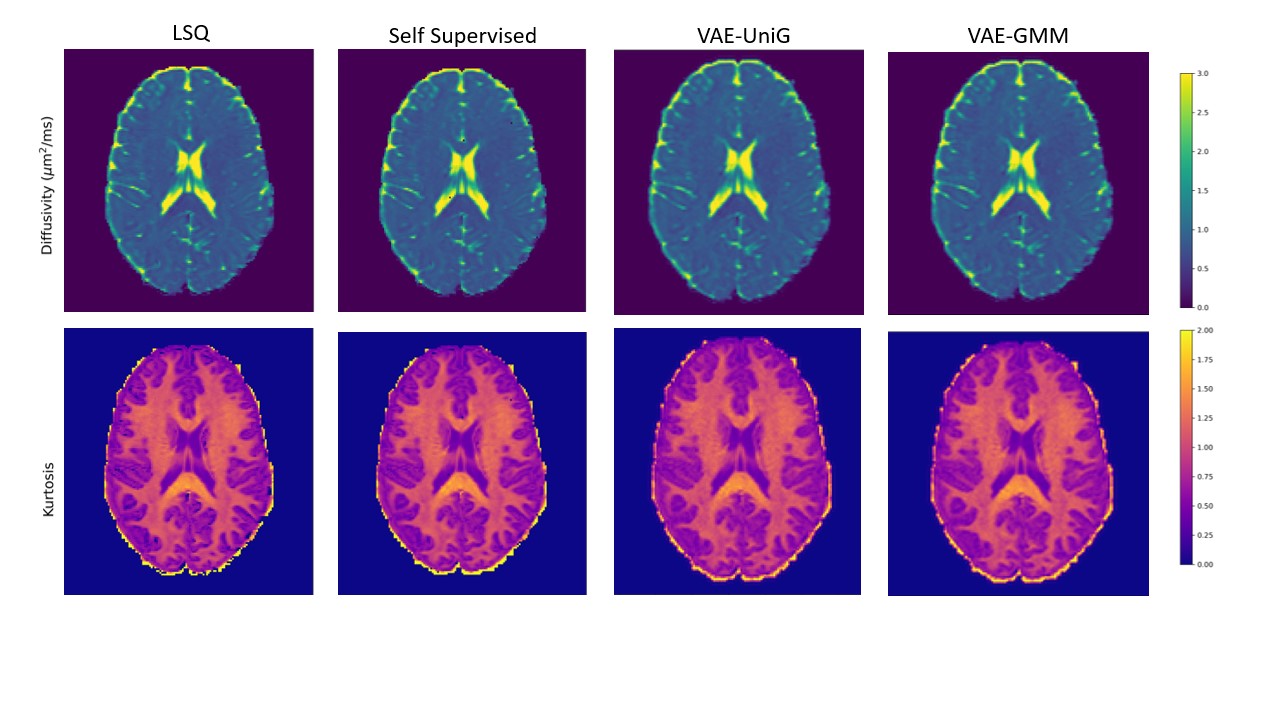}
    \end{center}
    \caption{Comparisons on MS-DKI fits on HCP dMRI subject. Our approach has less obvious improvements when the MRI model is relatively simple, but doesn't hallucinate spurious anatomical features.}
    \label{fig:ms_dki_HCP}
\end{figure}

\section{Results on the sensitivity of hyper parameters}
\begin{figure}[H]
\centering
\includegraphics[width=\textwidth]{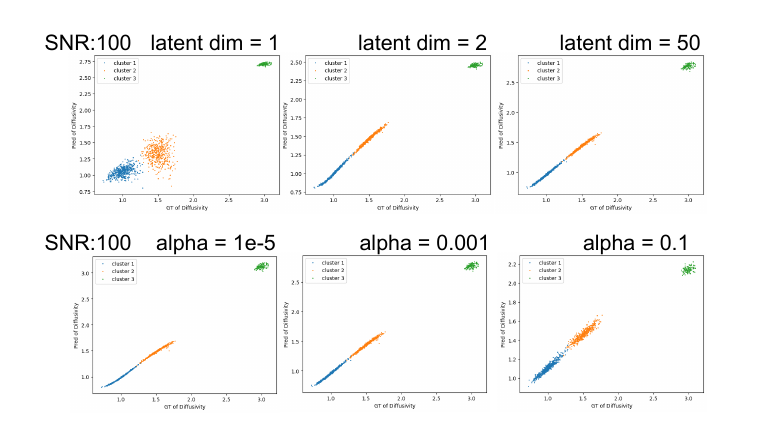}
\caption{Hyper-parameter sensitivities of our model with diffusivity results on simulated model using MSDKI, comparisons between self-supervised baseline and ours with Gaussian prior. X axis: ground truth of simulated diffusivity. Y axis: prediction of diffusivity. We observe that both latent dimension and kl loss strength have optimal values, but those values might be data dependant.}
\label{fig:simulated_hp}
\end{figure}

\section{Results on kurtosis estimation of MSDKI with VAE and simulated data}
\begin{figure}[!ht]
\centering
\includegraphics[width=\textwidth]{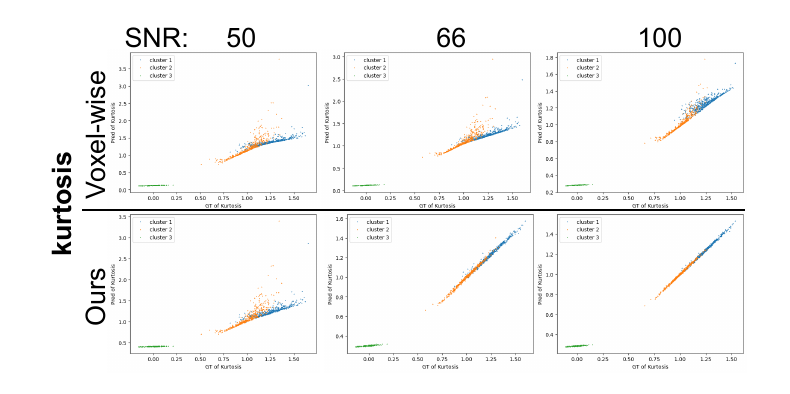}
\caption{Kurtosis results on simulated model using MSDKI, comparisons between self-supervised baseline and our VAE-UniG. X axis: ground truth of simulated diffusivity. Y axis: prediction of diffusivity. SNR: signal-noise-ratio. Blue: cluster 1. Orange: cluster 2. Green: cluster 3. Ours vastly outperforms the baseline in recovery of the ground truth.}
\label{fig:simulated_msdki_k}
\end{figure}

\end{document}